\documentclass[preprint,showpacs,preprintnumbers,amsmath,amssymb]{revtex4}

\usepackage{graphicx}
\usepackage{dcolumn}
\usepackage{bm}

\begin{document}

\title{\large Zero energy bound states in tunneling conductance spectra at the interface of an s-wave superconductor and a topological insulator\\
in NbN-$\rm Bi_2Se_3$-Au thin film junctions.}

\author{G. Koren}
\email{gkoren@physics.technion.ac.il} \affiliation{Physics
Department, Technion - Israel Institute of Technology Haifa,
32000, ISRAEL} \homepage{http://physics.technion.ac.il/~gkoren}

\author{T. Kirzhner}
\affiliation{Physics Department, Technion - Israel Institute of
Technology Haifa, 32000, ISRAEL}

\date{\today}
\def\bfig {\begin{figure}[tbhp] \centering}
\def\efig {\end{figure}}

\normalsize \baselineskip=8mm  \vspace{15mm}

\pacs{73.20.-r, 73.43.-f, 85.75.-d, 74.90.+n }

\begin{abstract}

Measurements of conductance spectra in a superconductor - topological insulator - normal metal thin film junctions of NbN-$\rm Bi_2Se_3$-Au are reported. Junctions with ex-situ and in-situ prepared $\rm NbN-Bi_2Se_3$ interfaces were studied. At low temperatures, all the ex-situ junctions showed coherence peaks in their conductance spectra, but imbedded robust zero bias conductance peaks were observed only in junctions with a metallic or a metal to insulator transition below $\rm T_c$ of the NbN electrode. The in-situ junctions which had about two orders of magnitude lower resistance at low temperatures, generally showed flat conductance spectra at low bias, with no coherence or broad Andreev peaks, since the critical current of the NbN electrode was reached first, at voltage bias below the energy gap of the superconductor. A weak zero bias conductance peak however, was observed in one of these junctions. We conclude that significant tunneling barriers, as in the ex-situ prepared junctions, are essential for the observation of coherence peaks and the zero energy bound states. The later seem to originate in the $\rm Bi_2Se_3$-NbN interface, as they are absent in reference Au-NbN junctions without the topological layer sandwiched in between.

\end{abstract}

\maketitle

\section{Introduction }
\normalsize \baselineskip=6mm  \vspace{6mm}

Topological insulators (TOI) became a hot topic in the past few years \cite{KaneRMP}. Ideally, they  should be bulk insulators with a finite energy gap, but with surface states (or edges states in 2D) which are gapless and conducting. These states had been observed in stoichiometric bulk chalcogenides such as $\rm Bi_2Se_3$ and $\rm Bi_2Te_3$ by angular resolved photoemission spectroscopy (ARPES) measurements \cite{ARPES}. In reality though, due to the high volatility of Se and Te, single crystals in general and thin films in particular are hole doped by vacancies of these atoms. As a result these materials become more conducting and even metallic. Further doping of these materials as was recently done in $Bi_2Se_3$ by copper intercalation, renders them superconducting with a transition temperature $\rm T_c$ in the range of 3-4 K \cite{Hor,Ando}. This could result in a new phase of matter which is a topological superconductor (TOS) whose hallmark signature would be the presence of Majorana fermions (MF). These MF  will lead to the appearance of a clear zero bias conductance peak (ZBCP) in conductance spectra of the TOS, which reflects their zero energy bound state nature \cite{Ando}. Sasaki \textit{et al}. had actually observed such a ZBCP in point contact measurements on superconducting $\rm Cu_xBi_2Se_3$ single crystals, and concluded from comprehensive theoretical considerations that they are due to MF which supports the TOS scenario \cite{Ando}. Similar kind of measurements were performed in parallel by our group, and the results support this conclusion \cite{Tal}. Another way to obtain superconductivity in a TOI is by means of the proximity effect with a known superconductor. Yang \textit{et al.} have used this method and observed a very narrow ZBCP ($\sim$0.05 mV) in junctions of Sn ($\rm T_c\sim 3.8$ K) and $\rm Bi_2Se_3$ single crystal flake at low temperatures \cite{LiLu}. In a previous study of our group robust ZBCPs were also observed in similar proximity systems where weakly superconducting $Bi$ induced superconductivity in $\rm Bi_2Se_3$ and $\rm Bi_2Te_2Se$ \cite{Koren}. In the present study we used the more robust NbN superconductor to induce superconductivity in $\rm Bi_2Se_3$, and investigated the transport properties of thin film junctions of $\rm Au-Bi_2Se_3-NbN$. Our results show that with significant tunneling barriers at the $\rm Bi_2Se_3-NbN$ interface, two types of robust ZBCP were observed which were absent in reference Au-NbN junctions. We conclude that these ZBCPs are due to proximity induced superconductivity in the surface states layer of the $\rm Bi_2Se_3$ film near the $\rm Bi_2Se_3-NbN$ interface.  \\

\begin{figure} \hspace{-20mm}
\includegraphics[height=9cm,width=11cm]{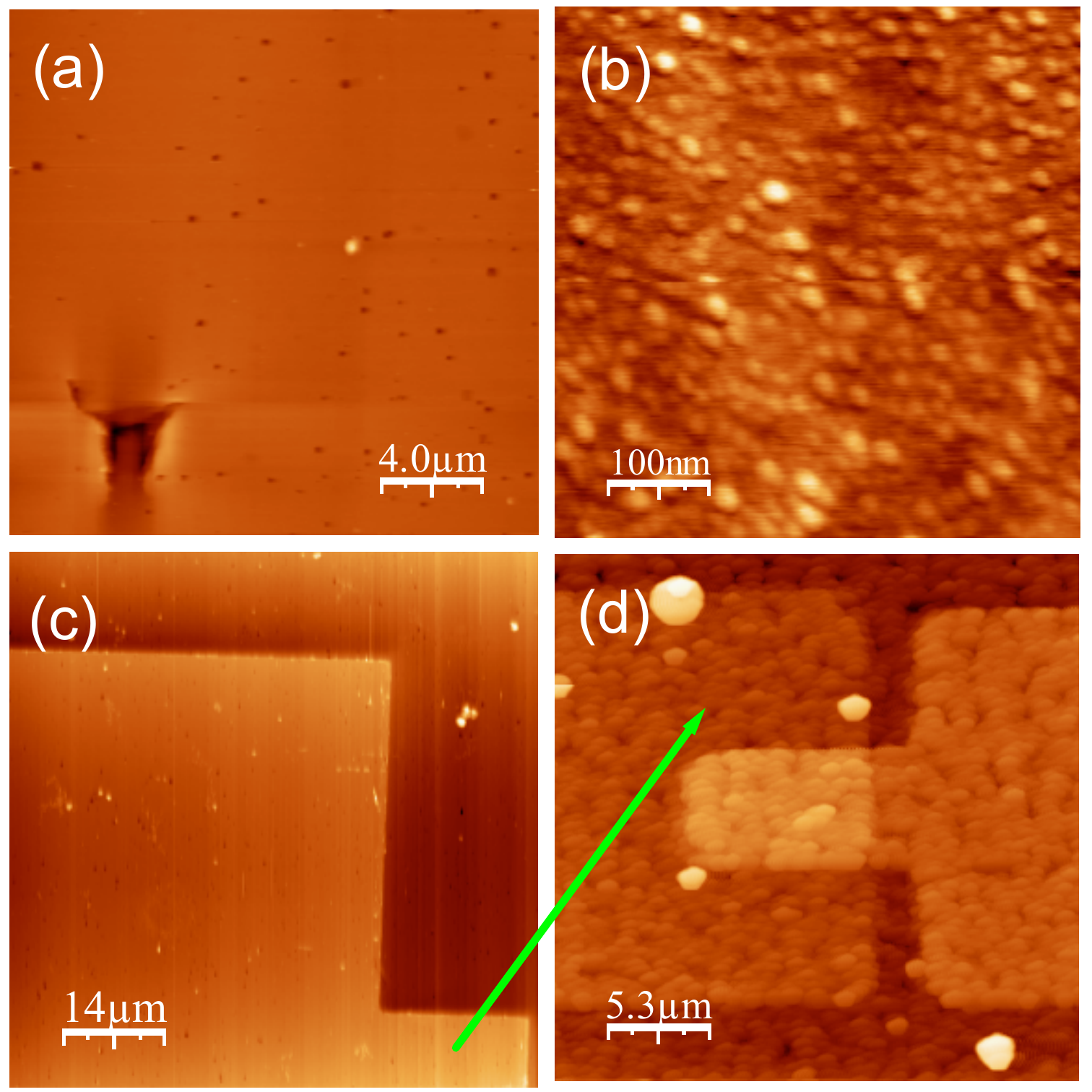}
\vspace{-0mm} \caption{\label{fig:epsart} (Color online) Atomic force microscope (AFM) images of a 70 nm thick NbN film before patterning (a) and (b), and after it (c) where a segment of the base electrode is seen. Image (d) shows an $\rm Au/Bi_2Se_3-NbN$ junction with a $\rm 10\times 5\,\mu m^2$ overlap area in the center (brightest). The arrow indicates the part of the base electrode of (c) as seen in (d).     }
\end{figure}

\section{Preparation and characterization of the films and junctions }
\normalsize \baselineskip=6mm  \vspace{6mm}

The NbN, Au and $\rm Bi_2Se_3$ thin films were prepared by laser ablation deposition from metallic Nb and Au targets and from a stoichiometric poly-crystalline $\rm Bi_2Se_3$ target. While the NbN film was deposited under 40-70 mTorr of $\rm N_2$ gas flow at 600 $^0$C heater block temperature, the Au and $\rm Bi_2Se_3$ films were deposited under vacuum at 150 and 400 $^0$C, respectively. High laser fluences were used for the deposition of NbN ($\sim 10\, J/cm^2$) and Au ($\sim 7 \,J/cm^2$), while a very low fluence was needed for the deposition of the $\rm Bi_2Se_3$ film ($\sim 1\, J/cm^2$) which also helped to reduce the Se loss. All films were deposited on (100) $\rm SrTiO_3$ wafers of $10\times10\, mm^2$ area. X-ray diffraction measurements of single layer NbN, and $\rm Bi_2Se_3$/NbN bilayer, showed that both grew with preferential crystallographic orientation. The NbN layer, mostly in the cubic phase with a-axis orientation and a=0.433 nm, while the $\rm Bi_2Se_3$ cap layer in the bilayer had the typical hexagonal structure with c-axis orientation normal to the wafer and c=2.88 nm. Figs. 1 (a) and 1 (b) show the smooth surface morphology of as deposited NbN films as seen by atomic force microscopy (AFM). Scattered small holes are observed in both Fig. 1 (a) (virgin surface) and 1 (c) (after patterning), while a rare big via-hole is seen in 1 (a). These holes might play an important role in the transport properties of junctions of the present study, as will be discussed later on.   \\

\begin{figure} \hspace{-20mm}
\includegraphics[height=9cm,width=11cm]{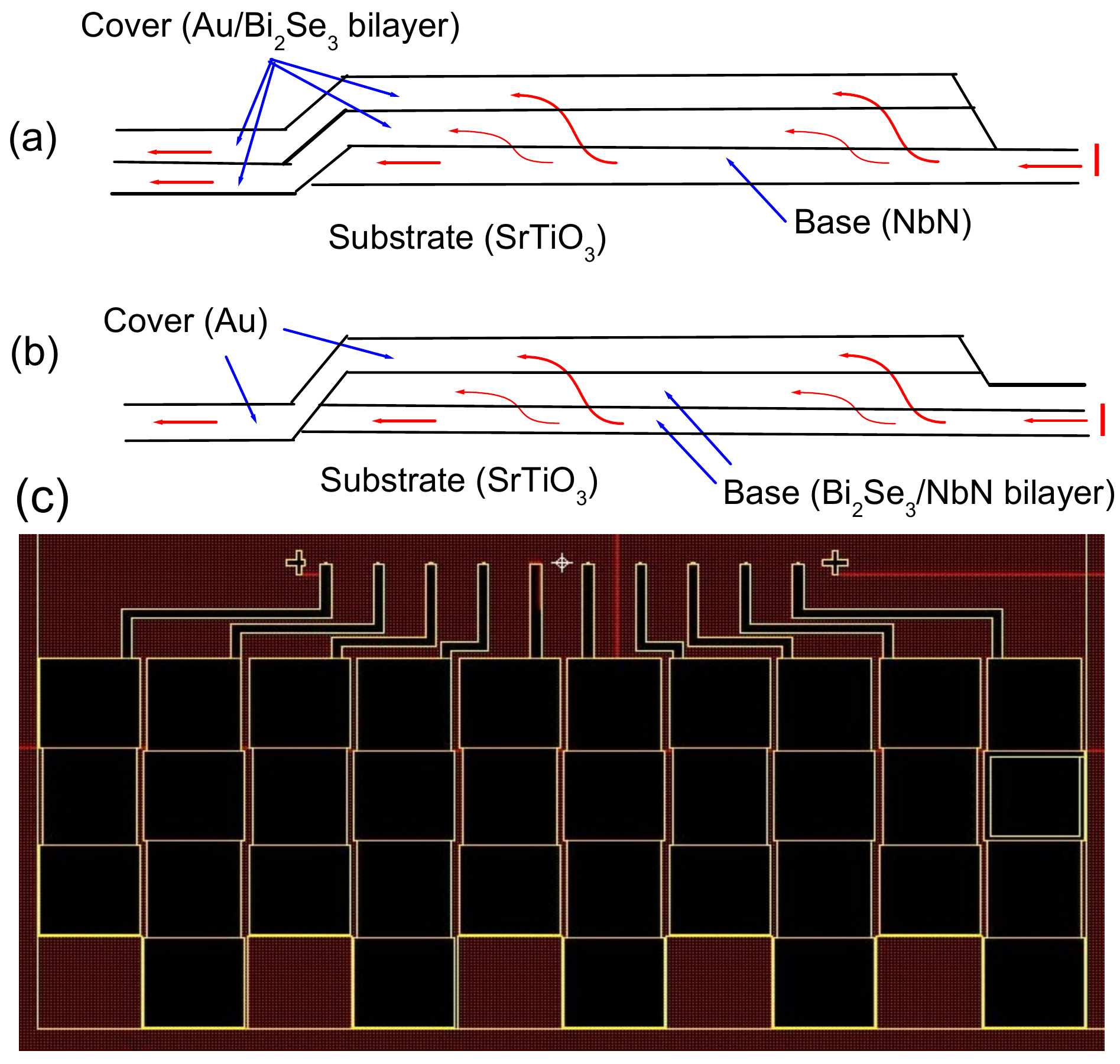}
\vspace{-0mm} \caption{\label{fig:epsart} (Color online) Panels (a) and (b) show schematic cross sections of the two types of junctions, with "ex-situ" and "in-situ" $\rm Bi_2Se_3$-NbN interfaces, respectively, where the current flow is indicated by red arrows. These plots are drawn not to scale, the trilayer segment represents the major overlap area of the junctions, which is typically one to two orders of magnitude larger than the ramp area (on the left). Panel (c) shows a schematic layout of the base electrode on half the wafer ($\rm 10\times 5\,mm^2$ area), with ten leads to the junctions (top) and twenty contact pads.  }
\end{figure}

\begin{figure} \hspace{-20mm}
\includegraphics[height=9cm,width=13cm]{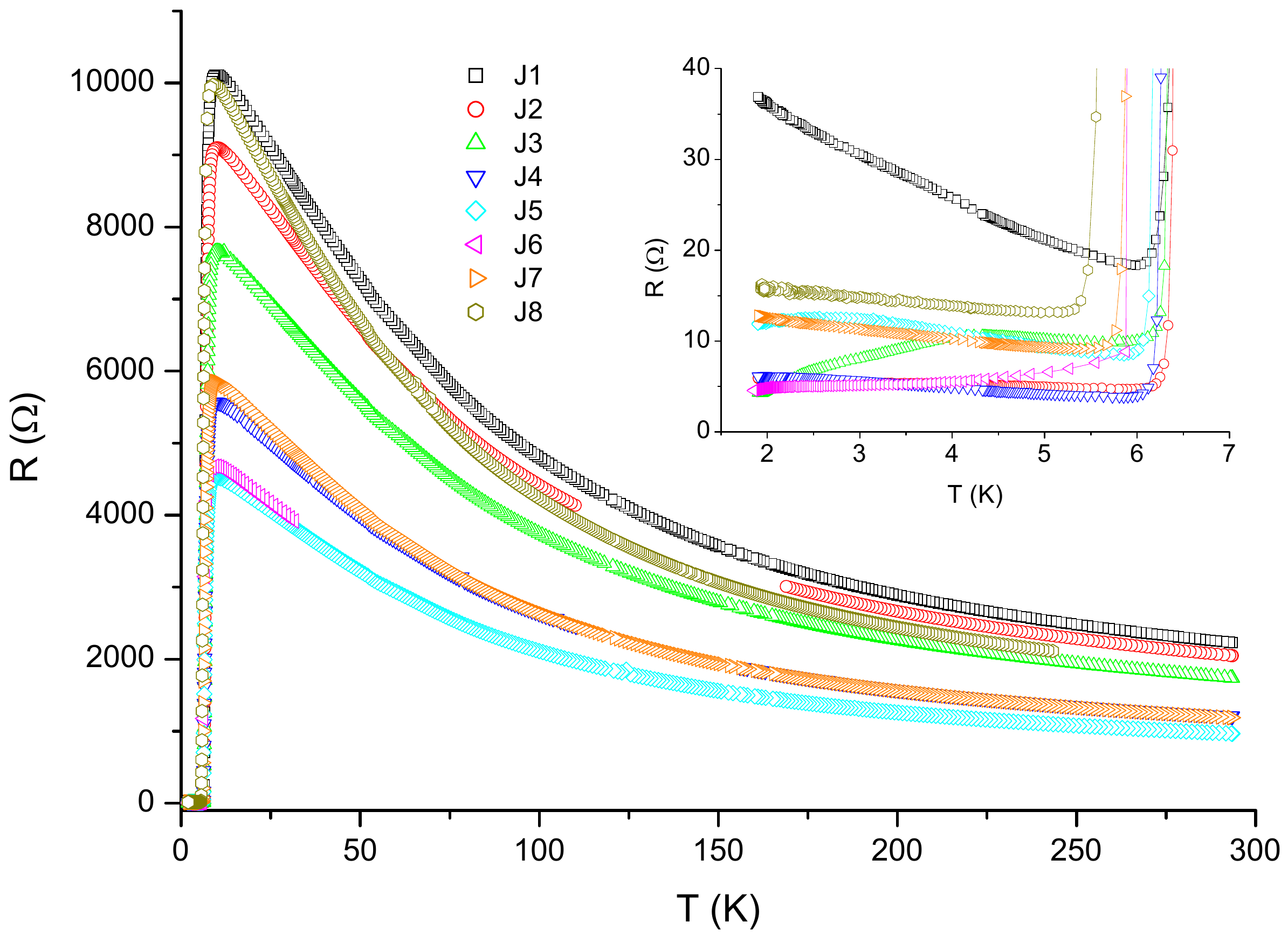}
\vspace{-0mm} \caption{\label{fig:epsart} (Color online) Resistance versus temperature of eight $\rm Au/Bi_2Se_3-NbN$ junctions of $\rm 150\times 100\,\mu m^2$ overlap area in the main panel, and a zoom-in on low temperatures in the inset. Of the other two junctions on the wafer, one was shorted and the other had a bad contact.  }
\end{figure}

The junctions were prepared from the different layers in a multi-step process using water-less PMMA resist, deep UV photolithography and Ar ion milling. This process which was successfully used in the past for the fabrication of ramp-type junctions of the cuprates \cite{Nesher}, was modified here by removing the insulating layer to have mostly "overlap" rather than "edge" (or ramp) junctions. Although the process was not fully optimized as of yet for the materials involved, we believe that our results on the proximity induced topological superconductivity here are sufficiently interesting, important and timely for reporting even at this early stage of the research. A typical top view of one junction is shown in Fig. 1 (d), and schematic cross sections are shown in Figs. 2 (a) and (b). First, a single layer of a 70 nm thick NbN film, or a bilayer of 100 nm thick $\rm Bi_2Se_3$ on 70 nm thick NbN layer were deposited. Ten base electrodes plus contact pads were then patterned on half the wafer as seen in Fig. 2 (c), followed by deposition of the cover electrode. In the case of the single layer NbN base, the cover electrode consisted of a 100 nm Au on 100 nm $\rm Bi_2Se_3$, while on the $\rm Bi_2Se_3$/NbN bilayer base electrode the cover was just a single 100 nm thick Au film. This led to geometrical differences as seen in Figs. 2 (a) and 2 (b), respectively, and to different $\rm Bi_2Se_3$-NbN interface properties. In the first case, an \textit{ex-situ} $\rm Bi_2Se_3$-NbN interface was formed, while in the second case, an \textit{in-situ} $\rm Bi_2Se_3$-NbN interface was involved. We shall therefore refer to the corresponding junctions as \textit{"ex-situ"} and \textit{"in-situ"} junctions.  The \textit{ex-situ} junctions were first produced with a large overlap area of the cover on the base electrode (about $\rm 150\times 100\, \mu m^2$), by the use of a shadow mask. Then after their characterization, further patterning of the gold layer of the cover electrode yielded ten junctions in the center of the wafer with $10\times 5\, \mu m^2$ overlap area each, as seen in Fig. 1 (d). In the \textit{in-situ} junctions as well as in reference NbN-Au junctions without the $\rm Bi_2Se_3$ layer, no shadow mask was used when depositing the gold cover electrode, and full photolithographic patterning yielded junctions similar to that of Fig. 1 (d), but with a smaller overlap area of $5\times 5\, \mu m^2$. An array of forty gold coated spring loaded spherical tips was used for the 4-probe measurements of the ten junctions. Although the junctions had the same geometry and size, their leads had different lengths as can be seen in Fig. 2 (c) and also from the different normal resistances in Fig. 3. \\

\section{Results and discussion}

\subsection{Large area \textit{ex-situ} junctions}

\begin{figure} \hspace{-20mm}
\includegraphics[height=9cm,width=13cm]{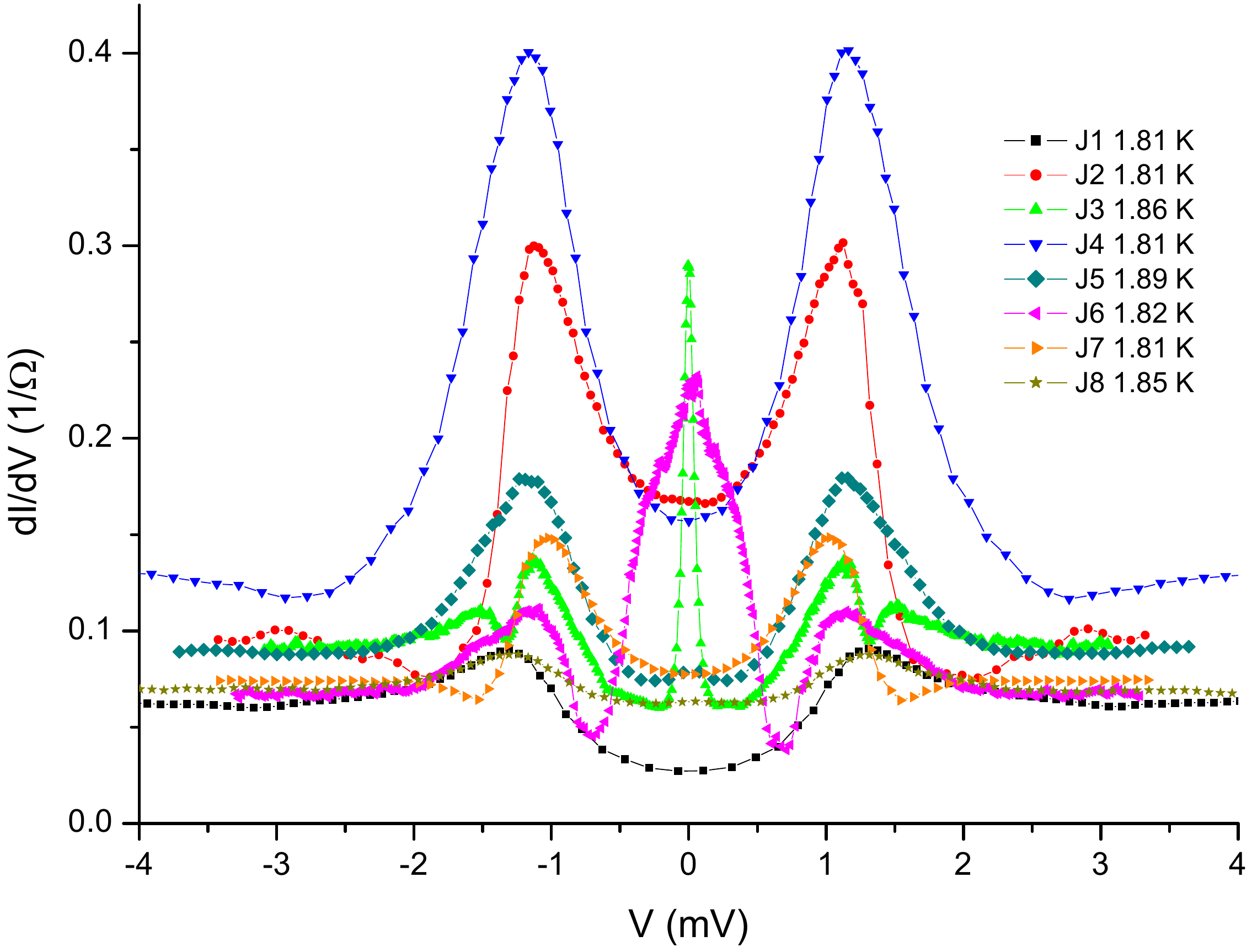}
\vspace{-0mm} \caption{\label{fig:epsart} (Color online) Conductance spectra at 1.8-1.9 K and zero field of the eight junction of Fig. 3. }
\end{figure}

Most of the interesting results of the present study were obtained on the large area (about $\rm 150\times 100\, \mu m^2$) \textit{ex-situ} junctions. Fig. 3 shows the resistance versus temperature of these junctions. The large and increasing normal resistance with decreasing temperature is due to the NbN base electrode film which was grown in excess nitrogen (70 mTorr). This lowered its' $T_c$ (onset at $\sim$8 K and offset at about 6 K), as can be seen in the inset to Fig. 3. We note that typically good laser ablated films have $T_c$ of about 15-16 K \cite{Treece}. Below $T_c$ offset the resistances typically ranged between 5-15 $\Omega$, which include mostly the junctions's resistance, but also a small contribution from the Au/$\rm Bi_2Se_3$ leads. The different behavior versus temperature of the different junctions, of slightly insulating, metallic, or with metal to insulator transition, are due to different tunneling barriers which were formed in the \textit{ex-situ} fabrication process \cite{Darlinski,Semenov}, but could also be affected by the scattered holes in the NbN layer as seen in Figs. 1 (a) and 1 (c).\\

\begin{figure} \hspace{-20mm}
\includegraphics[height=9cm,width=13cm]{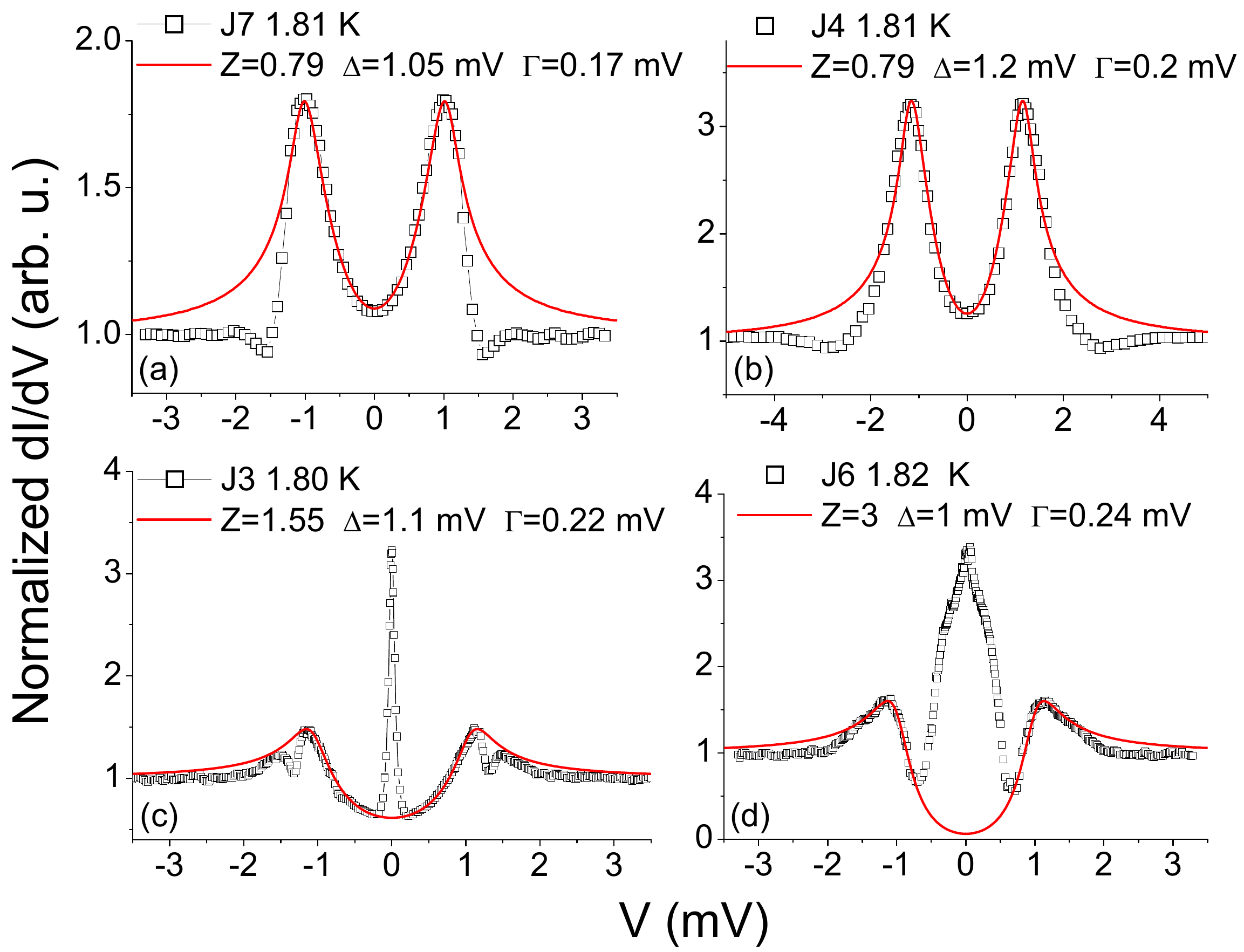}
\vspace{-0mm} \caption{\label{fig:epsart} (Color online) Normalized conductance spectra of four typical junctions of Fig. 4 (open squares), with BTK fits of an s-wave superconductor while ignoring the ZBCP features (red curves).  }
\end{figure}

Fig. 4 shows conductance spectra of all working junctions on the wafer where clear coherence peaks are observed in all junctions. The splitted or double coherence peak of J3 at $\sim$1.3 mV, as well as the sharp conductance drop of J2 at the same bias, is probably due to heating and critical current effects \cite{Sheet}. Fig. 5 shows the normalized conductance spectra of four representative junctions with different transparencies, together with Blonder - Tinkham - Klapwijk (BTK) fits for an s-wave superconductor \cite{BTK} while ignoring the ZBCPs. The resulting energy gaps $\Delta$ range between 1-1.2 mV, which is in-line with the relatively low $\rm T_c\sim 8$ K of these films, and in agreements with the classical BCS ratio $2\Delta/kT_c\simeq 3.5$ which yields $\Delta\simeq$1.2 meV.  The additional dip features in the spectra beyond the coherence peaks signify that the critical current in the superconducting NbN leads to the junctions was reached which involves heating and is typical of many junctions \cite{Sheet}. In addition to the coherence peaks, two robust ZBCPs are observed in two junctions on this wafer (J3 and J6), while weaker ones can be discerned in other junctions (J2, J5 and J8). These ZBCPs reflect the presence of zero energy bound states, which in no way can be originated in an s-wave superconductor as NbN is believed to be. So they should originate in the topological $\rm Bi_2Se_3$ layer near the interface with the NbN film. The proximity of this layer to the superconductor must render it superconducting by the proximity effect, and the induced superconductivity can be unconventional and give rise to the zero energy bound states. Theoretical calculations of tunneling conductance spectra in junctions of a topological superconductor and a normal metal showed robust ZBCPs for various odd-parity pair potentials \cite{Yamakage}. Similar Andreev bound states were observed in junctions of the cuprate superconductors along the nodes of the d-wave order parameter where it changes sign. They are also expected with other type or order parameters such as $p_x+ip_y$ as proposed for the topological superconductors \cite{Tanaka,p-wave,Hao}. Preliminary attempts by Yamakage and Tanaka to fit the present broad ZBCP of junction J6 of Fig. 5 (d) using their tunneling model with a single pair potential \cite{Yamakage}, yielded only a qualitative agreement with the data but not with its fine details. They made the conjecture that possibly the use of a combination of pair potentials as introduced by Fu and Berg \cite{FuBerg}, and allowed by the proximity effect, will yield a better fit to the data. Recently, topologically induced superconductivity in 1D semiconducting nano-wires have also shown ZBCPs which were field dependent in a way predicted by theory for Majorana fermions \cite{Kouwenhoven,Heibloom,Oreg}. We shall discuss the robust ZBCPs in more detail later on, but can already conclude that they originate in the proximity induced superconducting zone near the interface of the superconducting and topological layers.\\

\begin{figure} \hspace{-20mm}
\includegraphics[height=9cm,width=13cm]{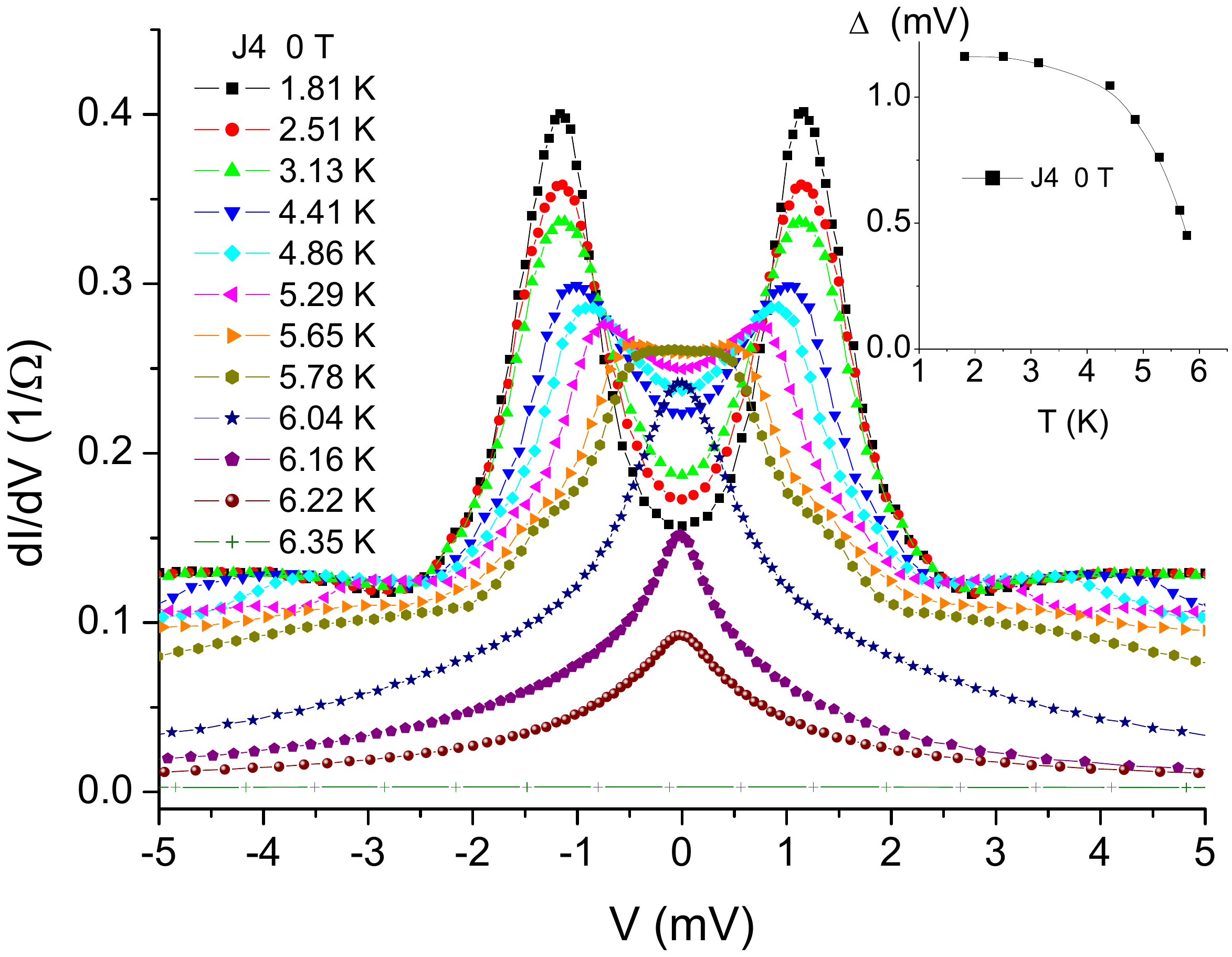}
\vspace{-0mm} \caption{\label{fig:epsart} (Color online) Conductance spectra at different temperatures of junction J4 of Fig. 5 (b). The inset shows the temperature dependence of the energy gap $\Delta$ as obtained from the coherence peaks' distance divided by 2.  }
\end{figure}

Next we present the temperature and field dependent conductance spectra of the junction J4 which has the largest coherence peaks but no clear ZBCP. These are shown in Figs. 6 and 7, where the coherence peak heights and corresponding energy gaps $\Delta$ are suppressed versus T and H as seen in these figures and their insets. For simplicity, the gap values were taken as the peak to peak distance divided by 2 which agreed quite well with the values of the BTK fits. The inset to Fig. 6 shows a typical BCS behavior of $\Delta$ versus T, while more data at higher fields is needed for the $\Delta$ versus H plot in the bottom inset of Fig. 7 in order to see a clear similar behavior. At higher temperatures, fields and voltage bias, the conductance goes down as heating effects start to play a role when either approaching $T_c$ with increasing temperature, or increasing the flux flow resistance at higher fields (lower conductance). The top inset of Fig. 7 shows a zoom-in on the low bias spectrum at 1.86 K. A small but clear ZBCP is now observed also in this junction, which is fully suppressed in a 6 T field, but fully recovered with no hysteresis, when this field is removed. We note that the width of this peak is similar to that of the narrow ZBCP of J3 in Figs. 4 and 5 (c).\\

\begin{figure} \hspace{-20mm}
\includegraphics[height=9cm,width=13cm]{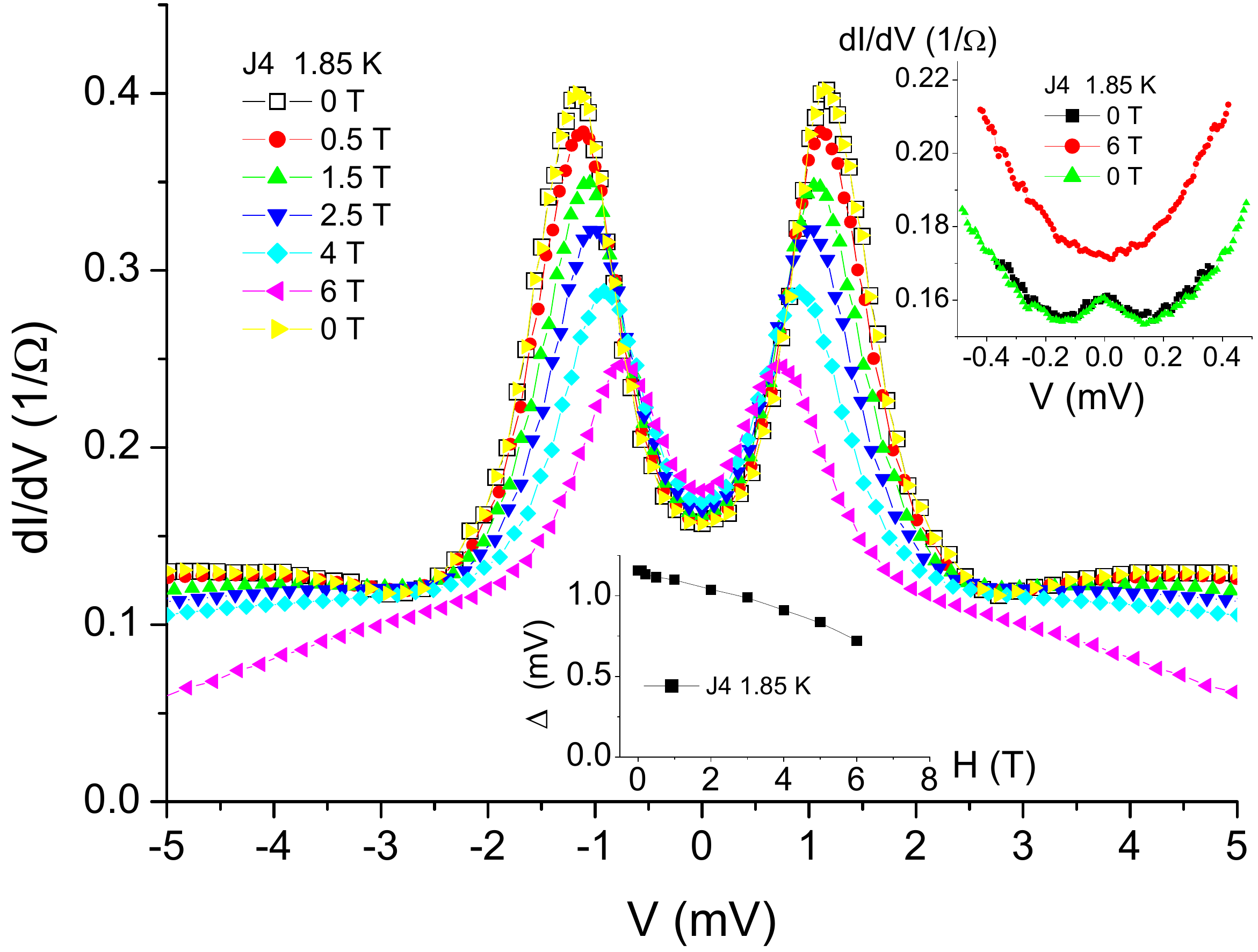}
\vspace{-0mm} \caption{\label{fig:epsart} (Color online) Conductance spectra at about 1.85 K under different magnetic fields normal to the wafer of the same J4 junction of Fig. 6. The top inset shows a small ZBCP at low bias which is fully suppressed under 6 T field. The bottom inset shows the gap energy $\Delta$ versus magnetic field. }
\end{figure}

\begin{figure} \hspace{-20mm}
\includegraphics[height=9cm,width=13cm]{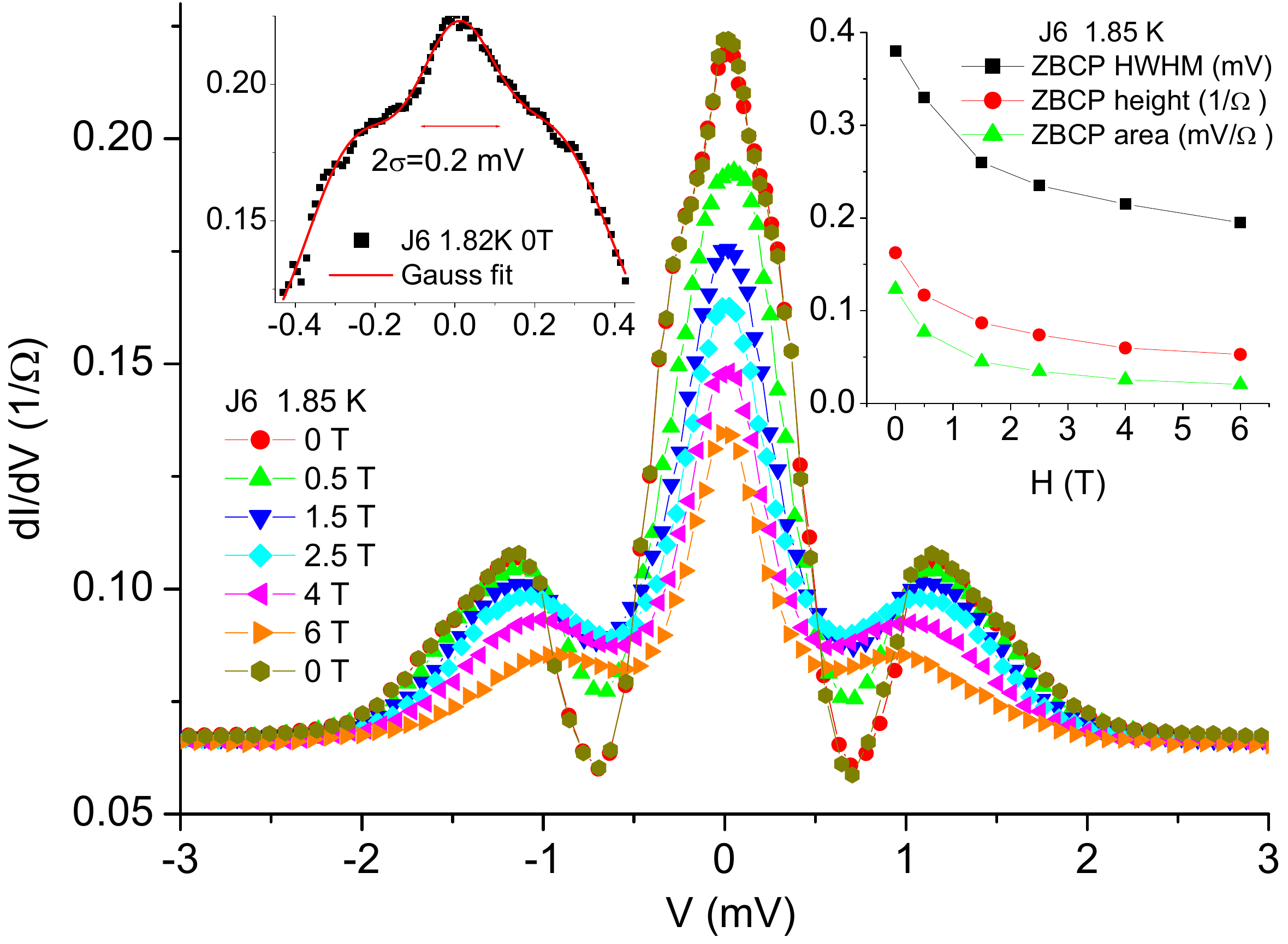}
\vspace{-0mm} \caption{\label{fig:epsart} (Color online) Conductance spectra at about 1.85 K under different fields of junction J6 of Fig. 5 (d). The right inset shows the magnetic field dependence of the ZBCP half width, height and area, all with respect to a base conductance at the minima on both sides of the ZBCP. The left inset shows a spectrum at 1.82 K with denser data in the low bias range, with a three peaks Gaussian fit.  }
\end{figure}

\begin{figure} \hspace{-20mm}
\includegraphics[height=9cm,width=13cm]{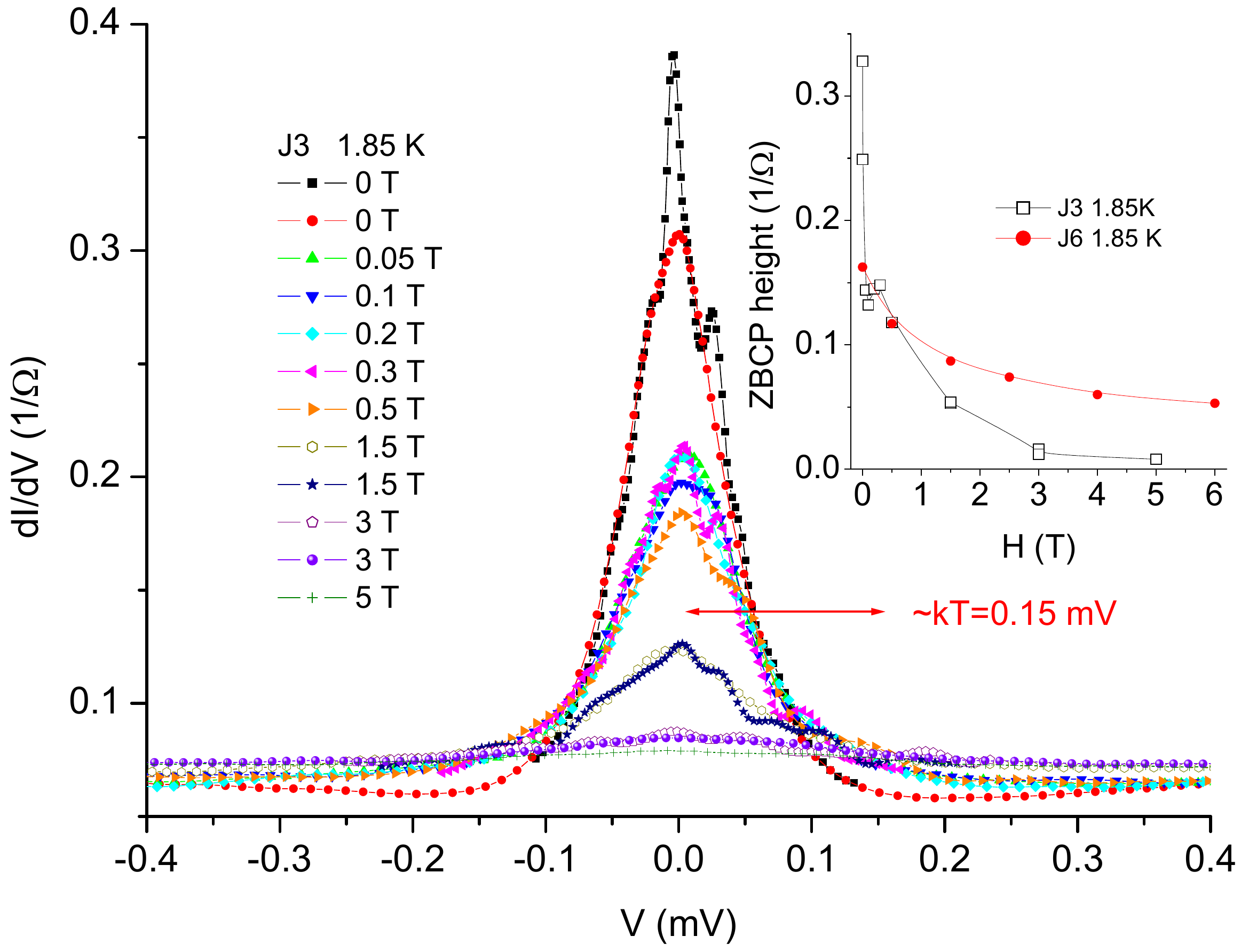}
\vspace{-0mm} \caption{\label{fig:epsart} (Color online) Low bias conductance spectra at about 1.85 K under different magnetic fields of junction J3 of Fig. 5 (c). The inset shows the field dependence of the ZBCP height of this junction, together with that of junction J6 of the inset to Fig. 8 for comparison.   }
\end{figure}

We shall now focus on the two types of robust ZBCPs observed in junctions J6 and J3 of Figs. 4, 5 (c) and 5 (d). Figs. 8 and 9 depict many conductance spectra of these two junctions at about 1.85 K under different magnetic fields. Since the field dependence of the coherence peaks of J6 in Fig. 8 is similar to what was observed before for J4 of Fig. 7, we shall not discuss it any more. The ZBCP heights in both Figs. 8 and 9 show suppression with increasing magnetic fields but the detailed field dependence is very different as can be seen in the inset to Fig. 9. While the broad ZBCP of J6 in Fig. 8 decays gradually with field, the narrow ZBCP of J3 in Fig. 9 has a very fast decay at low fields of up to $\sim$0.05 T, followed by a slower decay which is still faster than that of J6. So for this reason and for the clear difference in their width, the origin of the two ZBCPs must be different. The zero field conductance spectrum of J6 in Fig. 8 indicates that a narrower peak is superimposed on the broad ZBCP. In the left hand side inset to this figure, the low bias region is shown with higher resolution (denser data). The contribution of a clear narrow central peak is now observed, which has a 2$\sigma$ width of 0.2 mV as obtained by standard Gaussian fitting to 3 peaks. This width is similar to the half width at half maximum (HWHM) under 6 T field (right inset of Fig. 8), and also to the overall width of the narrow ZBCP of J3 in Fig. 9. We therefore propose that the the broad ZBCP of J6 contains two contributions: one of in-gap finite-bias bound states which seems to merge to a broader peak, and the other of a narrow peak due to zero bias bound states as in J3 of Fig. 9. It is clear from Fig. 9 that the overall width of the narrow ZBCP of J3 is field independent. This is due to the fact that this width is limited by thermal broadening at 1.85 K for which kT$\sim$0.15 meV. The seemingly narrower peak at 0 T (solid black rectangles in Fig. 9) is due to electronic measurement noise in this spectrum. Only measurements at lower temperatures, which are not available to us in the present system, will reveal the actual width of this ZBCP. Previous results in similar proximity systems showed overall ZBCP widths in the range of 0.03-0.05 mV \cite{LiLu,Kouwenhoven,Heibloom}, where the ZBCP themselves were attributed to Majorana fermions. Recently however, low bias bound states similar to the observed ZBCP were predicted to originate also in the topological trivial phase \cite{Brouwer}, so any experimental work that will shed light on this issue is important.  \\

\begin{figure} \hspace{-20mm}
\includegraphics[height=9cm,width=13cm]{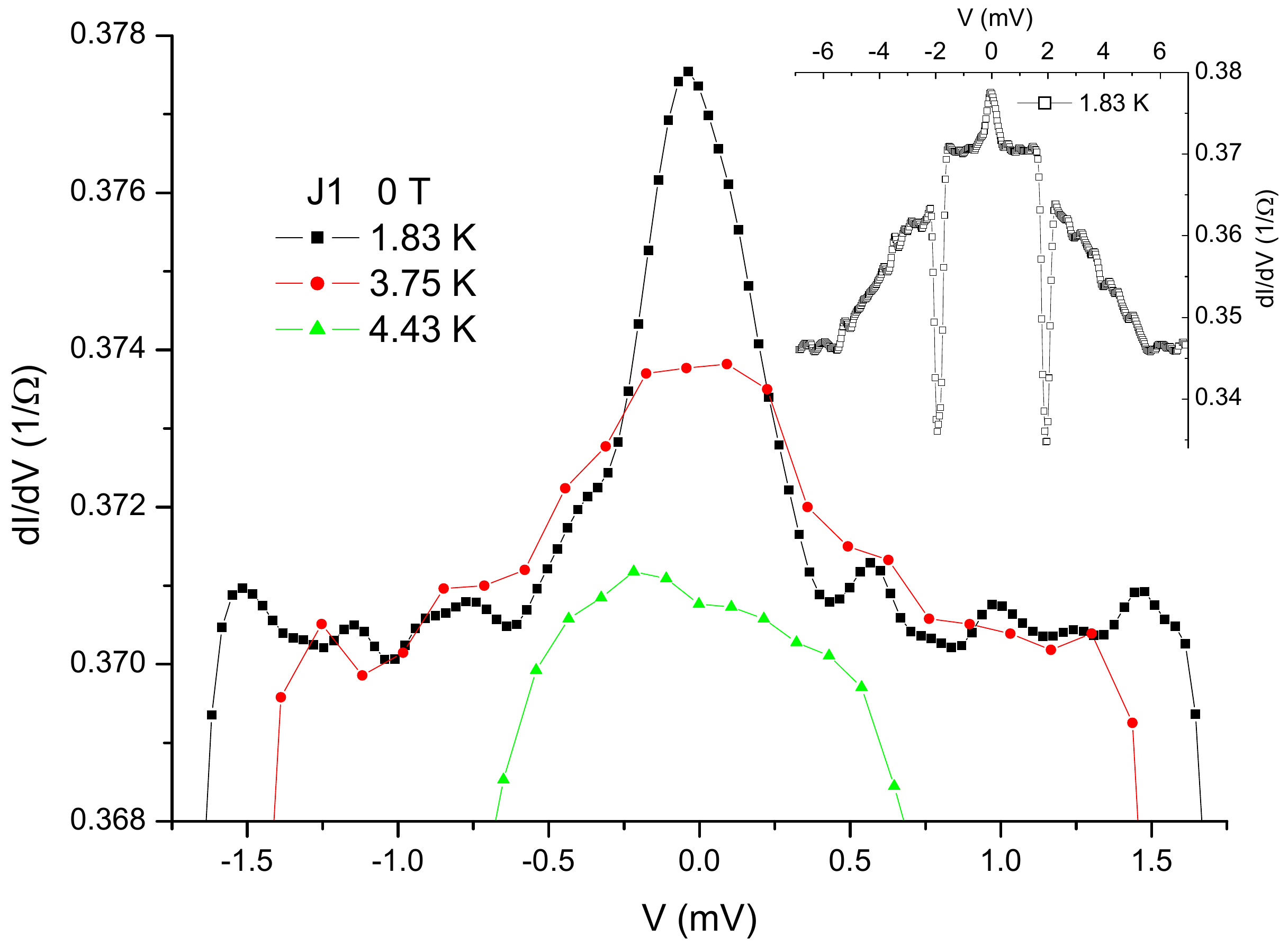}
\vspace{-0mm} \caption{\label{fig:epsart} (Color online) Conductance spectra under zero magnetic field and three different temperatures of a smaller junction on the same wafer of Fig. 3 after an additional milling step. All the gold layer and a 10 nm thick $Bi_2Se_3$ layer of the cover electrode overlapping the base electrode were milled off except for a $\rm 10\times 5\, \mu m^2$ area as seen in Fig. 1 (d). The inset shows the spectrum at 1.83 K over an extended bias range.   }
\end{figure}

\subsection{Small area \textit{ex-situ} junctions}

In an attempt to identify the part of the large area junctions which contributed to the observed results in Figs. 3-9, we made an additional patterning step  of the cover electrode on the same wafer to reduce the junctions area from $\rm 150\times 100\,\mu m^2$ to $\rm 10\times 5\,\mu m^2$. We milled away all the gold and 10 nm of the $Bi_2Se_3$ top layer of the large area junctions, except for the small overlap area of the desired small junctions. The final result of this procedure is shown in Fig. 1 (d). Although the junctions' area was reduced by a factor of 300, their overall resistance at low temperatures did not increase by this factor, but actually decreased by about an order of magnitude as compared to the original large junctions. The reason for this surprising result is that by milling directly on the $Bi_2Se_3$ layer, many Se vacancies were created, which contributed significantly to the overall junctions conductance in parallel to the contribution from the overlap area of the new small junctions. Nevertheless, a small ZBCP still survived in one junction as can be seen in the conductance spectra of Figs. 10 and 11. This ZBCP was suppressed by increasing temperature or increasing field at low temperature as before. Due to the small junctions resistance now ($\sim 2\,\Omega$), the critical current of the NbN lead to the junction was reached at lower bias as can be seen in the inset to Fig. 10 by the sharp dips at about $\pm$2 mV. These dips are due to local heating in the junction  \cite{Sheet}, which increase further with increasing bias. The ZBCP though is independent of these effects as it is measured at low bias (at low T and H). At higher temperatures and fields, the critical current is reached at lower bias, and this affects the side bands of the ZBCP as seen in Figs. 10 and 11. The ZBCP width is smaller than the previous broad one of the large area junction J6, but still larger than that of J3. This is probably due to the smaller signal and larger noise in the measurements, that prevents a more conclusive statement about width in this case. Concerning the fact that only one junctions on the wafer showed ZBCP now, we speculate that possibly the ZBCPs originate in the larger via holes in the base NbN electrode as in Fig. 1 (a), and because of their small abundance (none in Fig. 1 (c)), there are less ZBCPs in the small junctions. \\

\begin{figure} \hspace{-20mm}
\includegraphics[height=9cm,width=13cm]{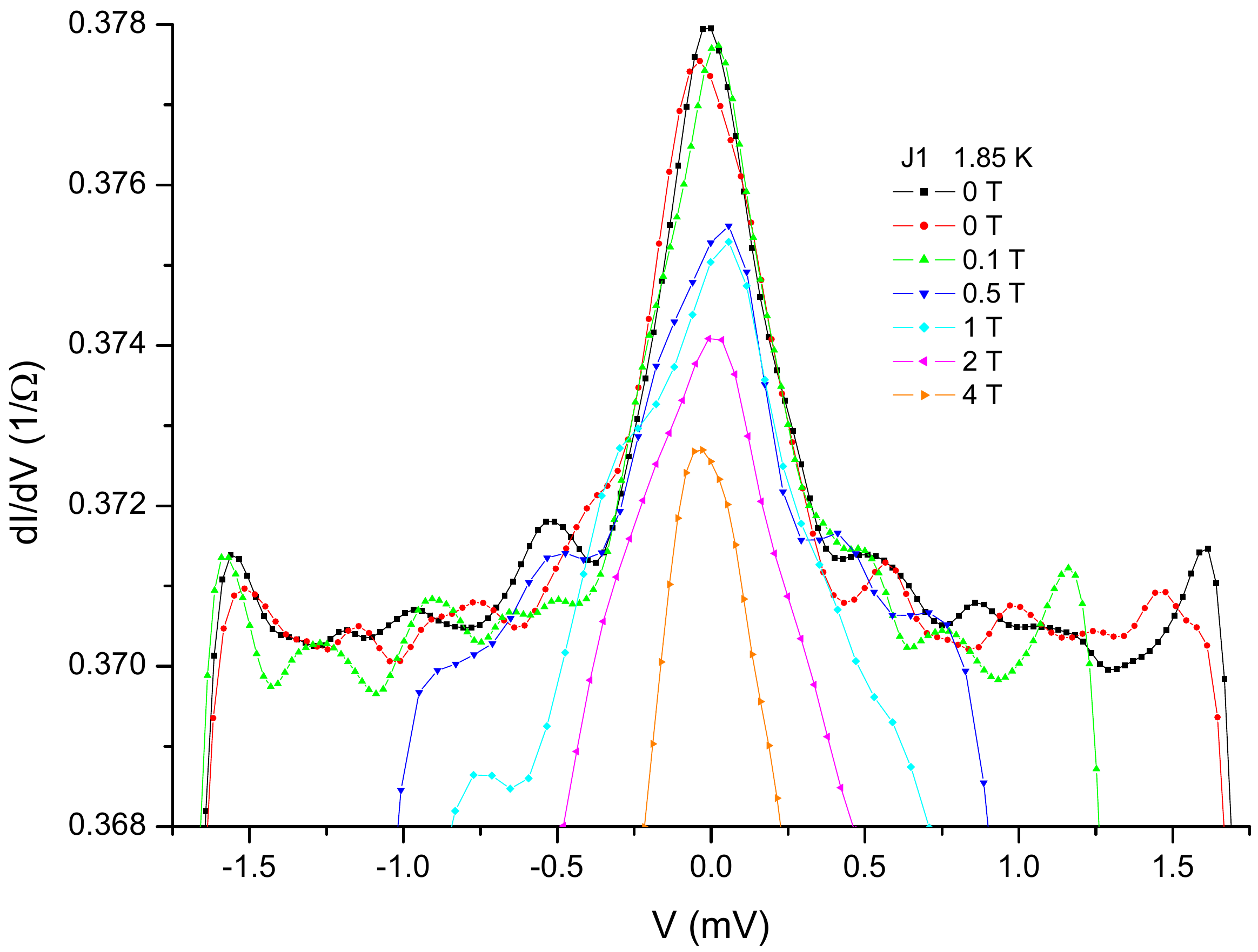}
\vspace{-0mm} \caption{\label{fig:epsart} (Color online) Conductance spectra of the same J1 junction of Fig. 10 at about 1.85 K and under different magnetic fields. }
\end{figure}

\subsection{Small area \textit{in-situ} junctions}

To check the effect of a high quality interface on the properties of our junctions, fully patterned small junctions of $\rm 5\times 5\,\mu m^2$ area were fabricated with an \textit{in-situ} deposited bilayer base electrode of $Bi_2Se_3$ on NbN, with a gold cover electrode and cross section as seen in Fig. 2 (b). This yielded junctions similar to that shown in Fig. 1 (d) but with half the overlap area. An AFM image of the as deposited bilayer is shown in the inset to Fig. 12. One can see that the top $Bi_2Se_3$ layer is well crystallized, but due to the lattice mismatch with the bottom cubic NbN layer, the hexagonal crystallites are laterally disordered (mosaic pattern). The resistance versus temperature of the resulting junctions is shown in Fig. 12. Due to the high conductance of the $Bi_2Se_3$ layer, the resistance of the leads and junctions is very low, less than 1$\Omega$ at room temperature. The $T_c$ of the NbN layer now is about 12.5 K, since it was deposited under 40 mTorr of nitrogen flow. The second transition at 8-9 K is attributed to the proximity induced superconductivity in the $Bi_2Se_3$ layer near the interface with the NbN. The residual low temperature resistance  of 60-110 m$\Omega$ is that of the junctions' interface, with a small contribution from the gold leads of the cover electrodes.\\

\begin{figure} \hspace{-20mm}
\includegraphics[height=9cm,width=13cm]{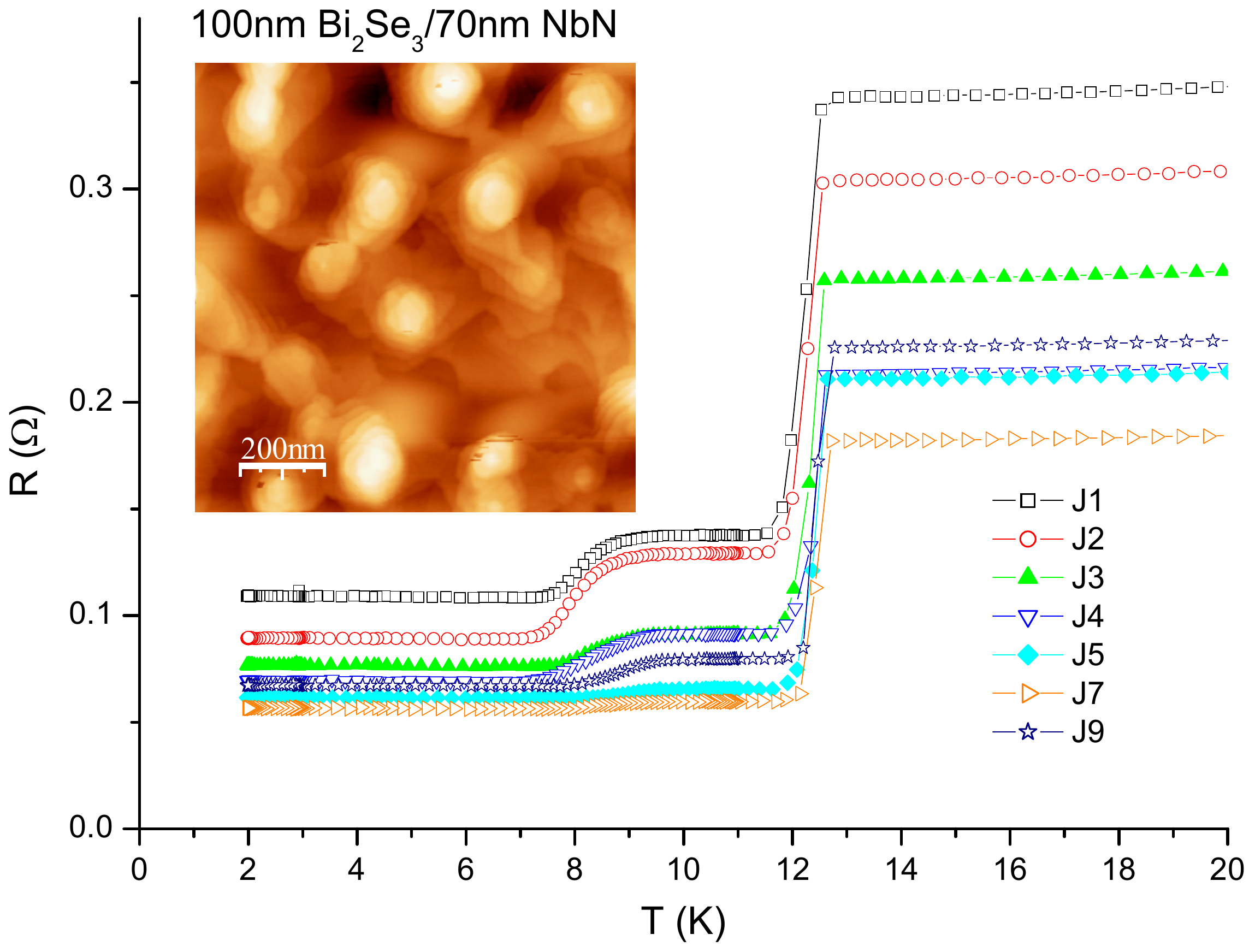}
\vspace{-0mm} \caption{\label{fig:epsart} (Color online) Resistance versus temperatures of seven junctions with an in-situ prepared interface of $\rm Bi_2Se_3$ on top of NbN base electrode and a gold cover electrode. The junctions' overlap area is $\rm 5\times 5\,\mu m^2$, and the geometry is as depicted in Fig. 2 (b) and similar to that seen in Fig. 1 (d). The high temperature data show metallic behavior versus T with R(300 K) ranging between 0.5-0.9 $\Omega$. The AFM image in the inset shows the the surface topography of the $\rm Bi_2Se_3$/NbN bilayer of the base electrode.  }
\end{figure}

\begin{figure} \hspace{-20mm}
\includegraphics[height=9cm,width=13cm]{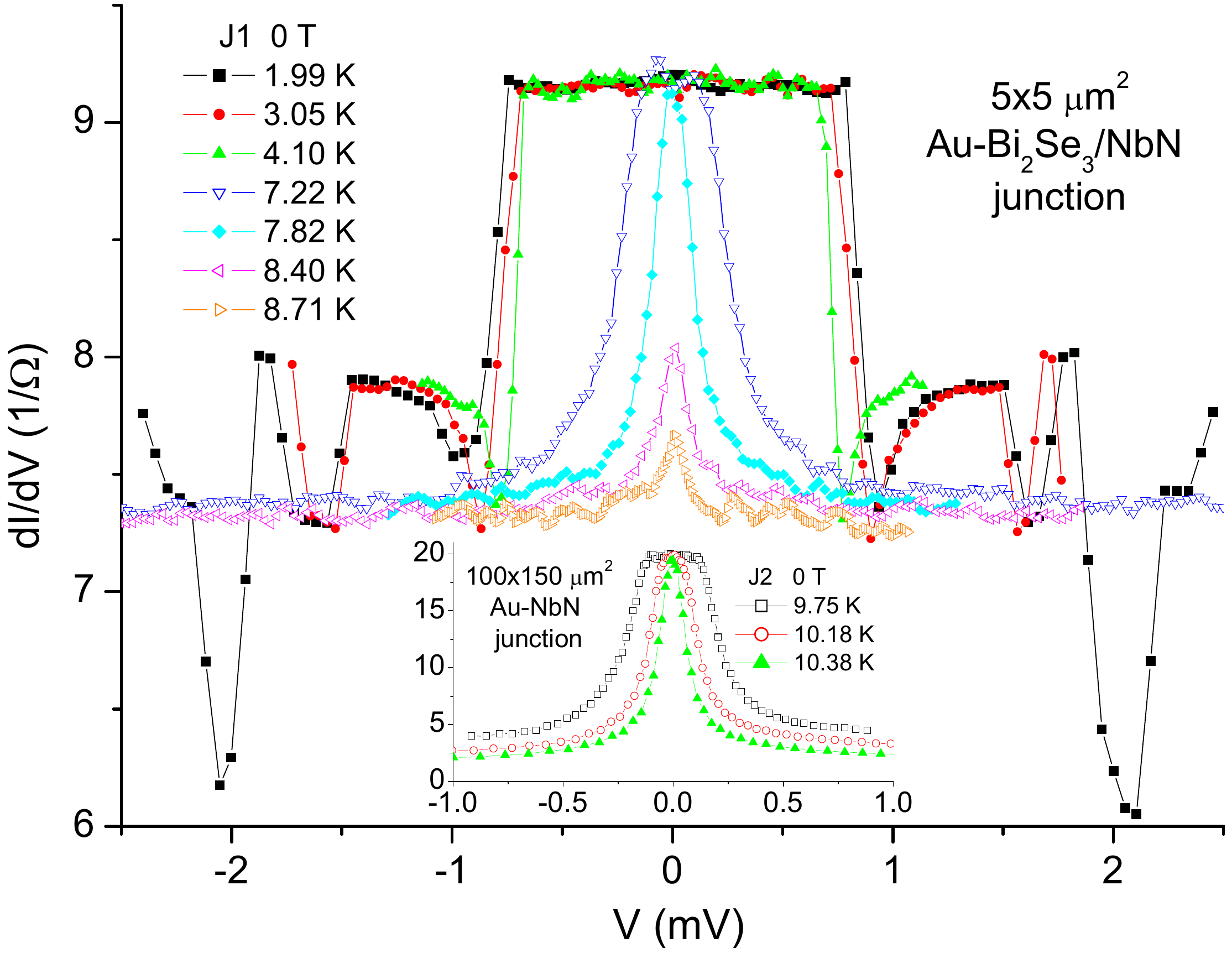}
\vspace{-0mm} \caption{\label{fig:epsart} (Color online) Conductance spectra at different temperatures and under zero field of junction J1 of Fig. 12. For comparison, the inset shows similar data in a reference Au-NbN junction without the  $\rm Bi_2Se_3$ layer. }
\end{figure}

Fig. 13 shows  typical conductance spectra of a junction on this wafer at different temperatures. As seen before in the spectrum of the inset to Fig. 10, due to the even lower junctions resistance, the critical current is reached at even lower bias voltages, and the resulting heating effects lead to the conductance dips. Therefore, coherence peaks at the energy gap bias, or a broad Andreev peak with width of twice the gap energy, now expected at about $\pm$1.5 mV due to the higher $T_c$, could not be observed. A ZBCP however could be observed, but nothing was found, possibly due to the rarity of the larger via holes in the NbN layer as explained before (see Fig. 1 (a)). With increasing temperature (and field, not shown) the critical current is reached at lower biases and seemingly emergent ZBCP is observed at about 8 K which coincides with the junction transition (see Fig. 12). This effect however, occurred also in a separately prepared reference wafer with NbN-Au junctions without the topological $Bi_2Se_3$ layer, as seen in the inset to Fig. 13. So this is not a real ZBCP, but a result of decreasing critical current as the superconducting $T_c$ is approached, and has nothing to do with the topological $Bi_2Se_3$. We thus conclude that a high quality $Bi_2Se_3$-NbN interface with low junction resistance is inappropriate for the observation of ZBCPs, and that a larger tunneling barrier, as in the case of the \textit{ex-situ} junctions, is essential for these observations. The null result of observing ZBCPs in the \textit{in-situ} junctions also rules out the possibility that the ZBCPs observed in the \textit{ex-situ} junctions are a spurious effect due to impurities, which might have been introduced into our junctions in the fabrication process.  \\

Finally we discuss an alternative scenario for the interpretation of our data. One may argue that the observed ZBCPs do not originate in zero energy bound states of the superconducting $Bi_2Se_3$ film, as there is no experimental proof of proximity induced superconductivity in this material in the present study. The second $T_c$ at 8-9 K in Fig. 12 which was attributed to the proximity effect, could be due to sample inhomogeneity. Furthermore, this second transition shows up in the \textit{in-situ} junctions, which do not have ZBCPs. On the other hand, \textit{ex-situ} junctions, where ZBCPs were observed, show a single $T_c$ around 6 K (Fig. 3), which originates in the NbN films. To refute this scenario, we first point out that the NbN sample inhomogeneity is small, as can be seen from the spread of $T_c(0)$ values of the electrodes in the \textit{ex-situ} junctions (5.4-6.3 K in Fig. 3), or even smaller in the \textit{in-situ} junctions (11.7-12.2 K in Fig. 12). Therefore, the second transition at 8-9 K in Fig. 12 can be attributed to the proximity effect and is not due to sample inhomogeneity. We also note that signatures of weak second transitions are observed in the \textit{ex-situ} junctions J3 and J5 of the inset to Fig. 3 at about 4.5 and 3 K, respectively, but these are too smeared to be considered as a reliable indication for the proximity effect. We therefore prepared another wafer recently of \textit{ex-situ} junctions that had NbN electrodes with $T_c$ of about 10 K, to check for the existence of a clear second transition and ZBCPs. We did observe a clear second $T_c$ at 5.5-7.5 K similar to that of Fig. 12 at 8-9 K, and a robust ZBCP similar to that of J3 of Fig. 9. Moreover, this ZBCP disappeared above 8 K while the gap depression of the density of states persisted up to 10 K, the transition of the electrodes.  We thus conclude that the second transition, at a few degrees below the transition of the electrodes, is in fact the smoking gun of the proximity induce superconductivity effect. In addition, reproducing the robust ZBCP on the new wafer, provides further support to our original interpretation of the data. These results will be published elsewhere. Currently, in collaboration with the group of Oded Millo, we are also looking for proximity effects and ZBCPs using scanning tunneling spectroscopy in $\rm Bi_2Se_3$ on NbN bilayers.\\

\section{Conclusions}

A study of proximity induced superconductivity in topological $Bi_2Se_3$ was carried out using Au-$Bi_2Se_3$-NbN thin film junctions. Various robust ZBCPs were observed in the junctions conductance spectra when the interface transparency was low. A narrow, temperature limited, ZBCP component was common to all the observed ZBCPs. We believe that this ZBCP has the same origin as the one observed by others in similar proximity systems and attributed to Majorana fermions.\\

{\em Acknowledgments:}  We greatly acknowledge Yamakage and Tanaka for their effort to fit our ZBCP data. This research was supported in part by the Israel Science Foundation, the joint German-Israeli DIP project and the Karl Stoll Chair in advanced materials at the Technion.\\

\bibliography{AndDepBib.bib}

\bibliography{apssamp}

\end{document}